\documentclass[aps,prl,preprint,superscriptaddress]{revtex4-1}

\usepackage{graphicx}
\usepackage{epstopdf}
\usepackage{hyperref}
\begin{document}


\title{Images of edge current in InAs/GaSb quantum wells}




\newcommand{\affSIMES}{\affiliation{Stanford Institute for Materials and Energy Sciences, SLAC National Accelerator Laboratory, Menlo Park, California 94025, USA}}

\newcommand{\affSUPhys}{\affiliation{Department of Physics, Stanford University, Stanford, California 94305, USA}}

\newcommand{\affSUAppPhys}{\affiliation{Department of Applied Physics, Stanford University, Stanford, California 94305, USA}}

\newcommand{\affRice}{\affiliation{Department of Physics and Astronomy, Rice University, Houston, Texas 77251-1892, USA}}

\author{Eric M. Spanton}
\affiliation{Stanford Institute for Materials and Energy Sciences, SLAC National Accelerator Laboratory, Menlo Park, California 94025, USA}
\affiliation{Department of Physics, Stanford University, Stanford, California 94305, USA}

\author{Katja C. Nowack}
\affiliation{Stanford Institute for Materials and Energy Sciences, SLAC National Accelerator Laboratory, Menlo Park, California 94025, USA}
\affiliation{Department of Applied Physics, Stanford University, Stanford, California 94305, USA}

\author{Lingjie Du}
\affiliation{Department of Physics and Astronomy, Rice University, Houston, Texas 77251-1892, USA}

\author{Gerard Sullivan}
\affiliation{Teledyne Scientific and Imaging, Thousand Oaks, CA 91630, USA}

\author{Rui-Rui Du}
\affiliation{Department of Physics and Astronomy, Rice University, Houston, Texas 77251-1892, USA}

\author{Kathryn A. Moler}
\affiliation{Stanford Institute for Materials and Energy Sciences, SLAC National Accelerator Laboratory, Menlo Park, California 94025, USA}
\affiliation{Department of Physics, Stanford University, Stanford, California 94305, USA}
\affiliation{Department of Applied Physics, Stanford University, Stanford, California 94305, USA}


\date{\today}

\begin{abstract}
Quantum spin Hall devices with edges much longer than several microns do not display ballistic transport: that is, their measured conductances are much less than $e^2/h$ per edge. We imaged edge currents in InAs/GaSb quantum wells with long edges and determined an effective edge resistance. Surprisingly, although the effective edge resistance is much greater than $h/e^2$, it is independent of temperature up to 30 K within experimental resolution. Known candidate scattering mechanisms do not explain our observation of an effective edge resistance that is large yet temperature-independent.
\end{abstract}


\maketitle

\section{}
 
A quantum spin Hall insulator (QSHI) is a two dimensional topological insulator that hosts counter-propagating spin-polarized edge states \cite{kane05prla, kane05prlb, bernevig06prl}. Elastic single-particle backscattering between counter-propagating edge states is not allowed because it violates time reversal symmetry \cite{qi11rmp}. Therefore the primary transport signature of the quantum spin Hall effect (QSHE) is ballistic conduction along the edges. Specifically, a single-mode ballistic channel will have a quantized conductance $e^2/h \approx (25.8 k\Omega )^{-1}$. The Landauer-B\"{u}ttiker formalism \cite{buttiker86prl} describes the conductance of multi-terminal devices with ballistic channels between contacts. Experimentally, conductances lower than those predicted by this model indicates that backscattering occurs along the edge. In the two known realizations of the QSHE, HgTe quantum wells \cite{bernevig06sci} and InAs/GaSb asymmetric quantum wells \cite{liu08prl}, sufficiently small samples show behavior that is generally consistent with ballistic edge channels \cite{konig07sci, roth09sci, du13arxiv, suzuki13prb}. Samples with longer edges have lower values of conductance \cite{konig07sci,knez11prl,du13arxiv,grabecki13prb,knez14prl}, implying the presence of scattering. InAs/GaSb offers a more accessible alternative to HgTe for exciting proposed devices. III-V growth is more widespread and its fabrication is more standard, and therefore understanding the disorder and scattering mechanisms of InAs/GaSb will be broadly important for investigating novel physics involving the QSHE.

In the absence of time-reversal-symmetry breaking, single-particle elastic backscattering is disallowed. Any observed backscattering should be explainable in terms of inelastic and/or multi-particle scattering. It remains unknown which scattering mechanisms are important in real materials. Candidates include magnetic impurities \cite{altshuler13prl,cheianov13prl} and nuclear spins \cite{lunde12prb, delmaestro13prb}. Various scenarios are based on disorder in the electric potential \cite{crepin12prb,schmidt12prl,lezmy12prb,vayrynen13prl,vayrynen14unpub}, which can occur due to impurities, dopants, or the gate dielectric, and may result in the formation of effective Kondo impurities \cite{maciejko09prl,tanaka11prl,altshuler13prl}. Although it may be possible to construct models with temperature-independent scattering over some range of temperature, inelastic processes should generally lead to a strong T-dependence. 

We previously showed, in HgTe quantum wells, that scanning superconducting quantum interference device (SQUID) magnetic flux images \cite{huber} can image edge currents and determine an effective edge resistance \cite{nowack13nmat}. Here, in InAs/GaSb, we used a four-terminal device to more accurately obtain the effective edge resistance at high temperatures in the presence of bulk conduction. We found that the effective edge resistance in InAs/GaSb is surprisingly independent of temperature, and that the two demonstrated QSHIs have surprisingly similar experimental signatures of scattering.

\begin{figure}
\includegraphics{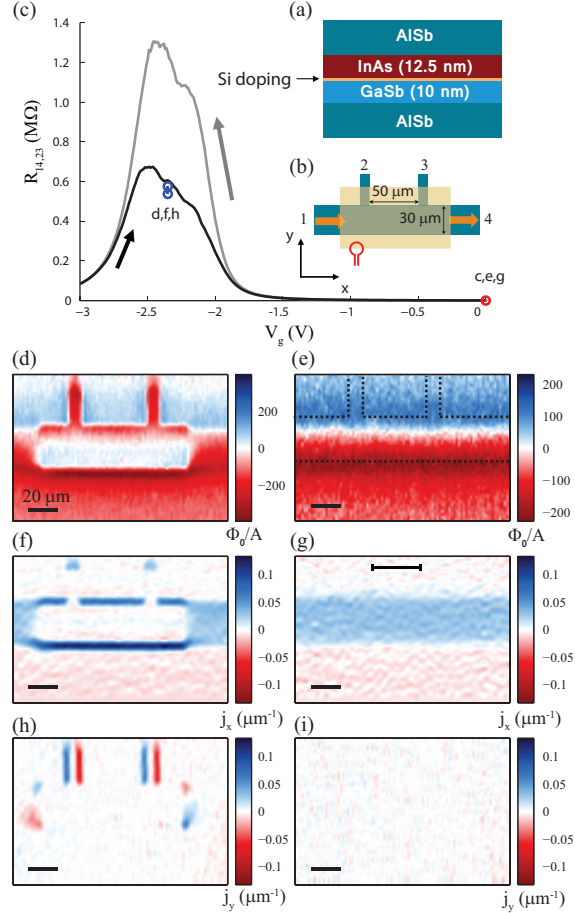}
\caption{\label{fig1} (color) Flux and current maps in a four-terminal device made from a Si-doped InAs/GaSb quantum well. (a) Schematic of the device. Si doping (shown in orange) suppresses residual bulk conductance in the gap. (b) Schematic of the measurement. Alternating current (orange arrows) flows from left to right on the positive part of the cycle. A voltage ($V_g$) applied to the front gate (yellow box) tunes the Fermi level. The SQUID's pickup loop (red circle) scans across the sample surface, with lock-in detection of the flux through the pickup loop from the out of plane magnetic field produced by the applied current. (c) Four-terminal resistance $R_{14,23} = V_{23}/I_{14}$ as a function of $V_g$, showing both the upwards (black) and downwards (gray) gate sweeps. We measured resistance before and after each image (blue and red circles).$ R_{14,23} $ is maximized when the chemical potential is tuned into the gap. (d,e) Flux images for the sample tuned into (d) the bulk gap, $V_g$ = -2.35 V, and (e) the n-type regime, $V_g$ = 0 V. (f-i) Reconstructed horizontal ($j_x$) and vertical ($j_y$) 2D current densities, showing that the current flows on the edges in the bulk gap and uniformly outside the gap. The black bracket in (g) indicates the region of averaging to obtain FIG.~\ref{fig2}b and dashed lines in (e) indicate the approximate geometry of the sample. The zero of flux in (e) is not in the center of the device due to the asymmetric geometry of our SQUID's pickup loop.}
\end{figure}

We studied several devices made from two InAs/GaSb quantum wells, one without doping and one with Silicon doping at the InAs/GaSb interface. In undoped InAs/GaSb quantum wells, residual bulk conductivity complicated the initial studies of the QSHE \cite{naveh01epl,knez11prl}. We imaged current flow and confirmed the coexistence of bulk conduction and enhanced edge conduction in a device made from an undoped quantum well \cite{som}. Silicon doping at the interface \cite{du13arxiv}, Beryllium doping in the barrier layer \cite{suzuki13prb}, and use of a Gallium source with charge-neutral impurities \cite{charpentier13arxiv} all reduce the residual bulk conductivity. Here we investigated a device made from a wafer (FIG.~\ref{fig1}a) with $\sim 10^{11}$ $cm^{-2}$ Si dopants at the InAs/GaSb interface and layer thicknesses that are predicted to result in an inverted band structure exhibiting the QSHE \cite{liu08prl}. Observation of dissipative non-local transport at high fields in InAs/GaSb suggests inversion of the lowest Landau levels, which is consistent with the inverted band structure necessary for the QSHE \cite{nichele14prl}. The growth of the Si-doped wafer is described in Ref. \cite{du13arxiv}. FIG.~\ref{fig1}b shows a schematic of the device. The lengths of the edges of our device ($>50$ $\mu m$) are much larger than the phase coherence length observed in similar samples (4.2 $\mu m$) \cite{du13arxiv}. Such edges exhibit backscattering and their resistance scales with length \cite{du13arxiv}. 

To determine the 4-terminal resistance ($R_{14,23}$)  of the sample, we applied current from contacts 1 to 4, and measured the voltage between contacts 2 and 3. Each reported resistance was either measured at a root mean square (rms) current of 10 nA at quasi DC frequencies ($<5$ Hz) or extracted from fitting full I-V characteristics ($<10$ nA). $R_{14,23}$ at zero applied front gate voltage ($V_g=0 V$) is ~10 $k\Omega$. Using the front gate, we depleted n-type carriers by applying a negative gate voltage. FIG.~\ref{fig1}c shows $R_{14,23}$ as a function $V_g$. We observe a maximum in $R_{14,23}$ at $V_g = -2.35 V$, indicating that we have tuned the chemical potential into the device’s insulating gap. The maximum value of $R_{14,23} >> h/(4e^2)$ indicates that backscattering occurs along the edges. At more negative voltages, $R_{14,23}$ decreases again, indicating that the chemical potential lies in the valence band and that the majority of carriers are p-type.

As is often the case in gated devices, $R_{14,23}$ depends on the gate voltage history. The device was consistently more resistive on downward sweeps of the gate. Such history dependence implies that repeated sweeps, as well as temporal drift, likely give different realizations of the disorder potential.

To image current, we applied an AC current with a nominal rms amplitude of 150 nA and used lock-in techniques to measure the resulting flux through the SQUID's 3-$\mu m$-diameter pickup loop (shown schematically in FIG.~\ref{fig1}b). We corrected the images and profiles presented in this letter for phase shift and attenuation from unintentional RC filtering. 

FIG.~\ref{fig1} shows two images of magnetic flux produced by current in the InAs/GaSb device, contrasting the cases where the chemical potential was tuned into the conduction band ($V_g$ = 0 V) and into the gap ($V_g$ = -2.35 V). At $V_g$ = 0 V, the magnetic flux varied smoothly and monotonically across the device (FIG.~\ref{fig1}e), indicating that current flowed uniformly inside the sample. When the device was tuned near its resistance peak ($V_g$ = -2.35 V), the flux had sharp features centered on the edges of the device, signifying that current flowed along the edges of the sample (FIG.~\ref{fig1}d). 

To better visualize the current, we used Fourier techniques \cite{roth89jap,nowack13nmat} to extract the 2D current density from each flux image. The resulting current images confirm that in the conduction band, the current distributed uniformly throughout the device (FIG.~\ref{fig1}g,i). In the gap, however, the current flowed almost entirely along the edges under the front gate (FIG.~\ref{fig1} f,h), a signature of the QSHE. Edge currents are particularly illustrated in the vertical leads (FIG.~\ref{fig1}h), in which current flowed along the lead until it reached an un-gated part, where the current crossed and returned along the opposite edge of the lead. The qualitative features of the images did not depend on gate voltage history.

Observation of ballistic conduction in $1 \mu m$ wide devices \cite{du13arxiv} sets an upper limit of $\sim 500 nm$ on the width of the edges, below our spatial resolution. The geometry of the SQUID's $3 \mu m$-diameter-pickup loop, the height above the sample ($\sim 1.5 \mu m$), and the current inversion all limited our spatial resolution and determined the apparent width of the edge conduction. The expected signal from spin-polarization of electrons is below our experimental sensitivity and is complicated by the presence of the much larger magnetic fields from current flow \cite{som}.

In the supplementary information, we checked the possible impact of nonlinearity on our results by measuring I-V characteristics and taking images at different current amplitudes \cite{som}.

\begin{figure}
\includegraphics[width=4.0in]{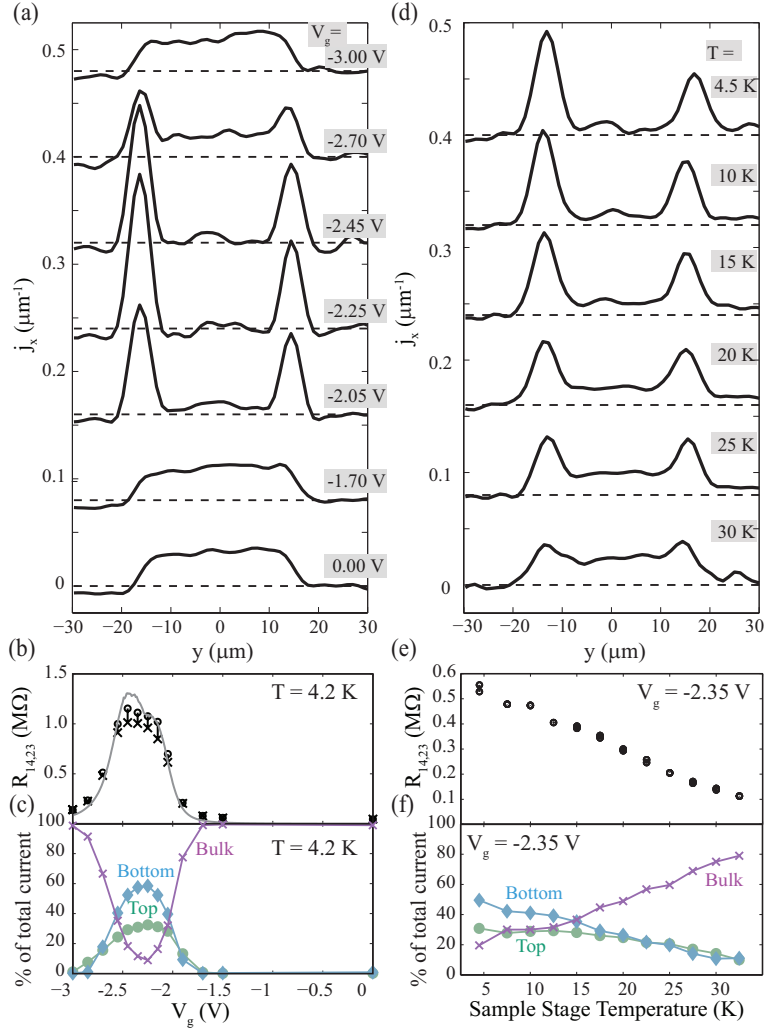}
\caption{\label{fig2} (color online) Analysis of current flowing along the edges as a function of $V_g$ (a-c) and temperature (d-f) (a) Selected profiles of the x-component of the current density show the evolution from bulk-dominated to edge-dominated transport, offset for clarity. The zero of each profile is indicated by the dashed line. Profiles were averaged over the region between the contacts, as shown in Fig. 1g. (b) Resistance vs. $V_g$ in a downward gate voltage sweep before imaging the current (gray), and immediately before (o) and after (x) each image in a subsequent sweep. (c) The fitted percentage of current flowing in the top edge (green circles), bottom edge (blue diamonds) and bulk (purple x’s) as a function of $V_g$. (d) Profiles of the x-component of the current density at selected temperatures, showing more bulk conductivity at higher temperatures, and the presence of edge states up to 30 K. The profiles are offset for clarity and the zero of each profile is indicated by a dashed line. (e) $R_{14,23}$ of the device as a function of temperature. (f) Fitted percentage of current flowing in the top (green circles), bottom (blue diamonds), and bulk (purple x’s). For (a) and (d), the origin in y is defined with respect to the center of the device. }
\end{figure}

To understand the evolution of current with gate, we imaged current at a series of gate voltages (FIG.~\ref{fig2}). The gate was swept downward and the resistance was recorded before and after each scan FIG~\ref{fig2}b. In FIG.~\ref{fig2}a we present selected profiles of the x-component of the current density as a function of gate voltage. The area over which the profiles were averaged is indicated in FIG.~\ref{fig1}g. We fit flux profiles, as described in \cite{nowack13nmat}, to quantify the amount of current flowing in the top edge, in the bottom edge, and homogeneously through the bulk (FIG.~\ref{fig2}c). As the bulk conductance increased, the amount of edge current decreased. The top edge current was more strongly influenced than the bottom edge current, because the bulk conductance provided an alternative path across the top leads, effectively decreasing the length of the current path along the top edge. 

At $V_g$ = -2.35 V there was little bulk conductance and the current traced out the entire top edge. Under these conditions, the top current was approximately half of the bottom edge current, consistent with an edge resistance that scales with length.

Next we studied how the current distributed as a function of temperature while fixing $V_g$ at   -2.35 V. The resistance peak remained at the same gate voltage value over the range of temperatures measured. We measured flux profiles along y at the center of the device between 4.5 K and 32.5 K. We extracted a 2D current density from a single flux profile (FIG.~\ref{fig2}d). The total resistance dropped with increasing temperature (Fig.~\ref{fig2}e) as more current flowed in the bulk of the sample, indicating that the bulk's conductivity is  increasing relative to the edges. We fitted the flux profiles as a function of temperature as done above (FIG.~\ref{fig2}f).

\begin{figure}
\includegraphics[width=4.0in]{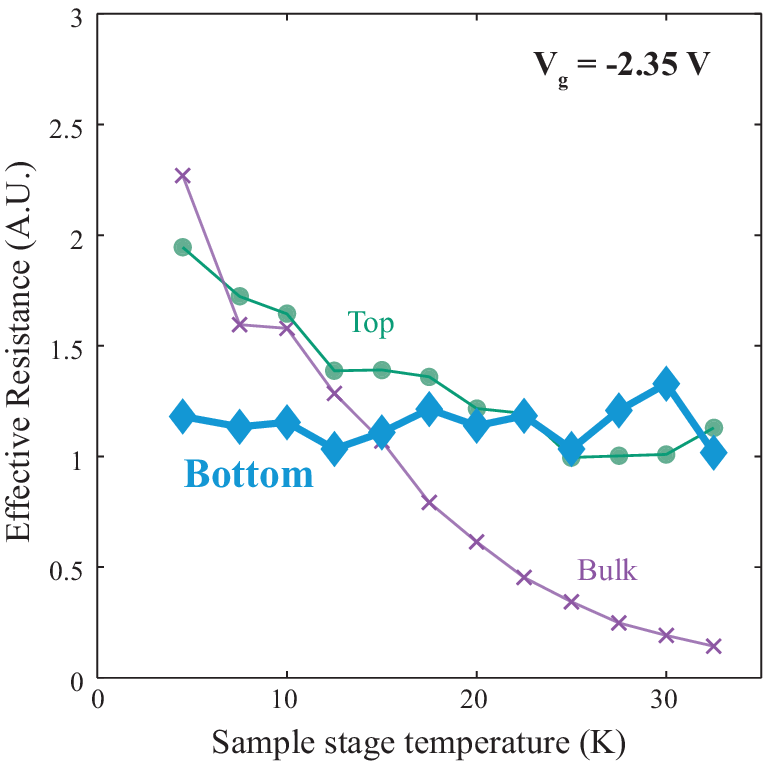}
\caption{\label{fig3} Effective resistance of the bulk and edges as a function of temperature in the quantum spin Hall state. The effective resistance of the bulk varies strongly with temperature. The effective resistance of the top edge is reduced by a factor of two, consistent with shorting of the top contacts by the bulk, rather than a change in the resistance per unit length (see text). The effective resistance of the bottom edge remains constant with temperature, which known scattering theories do not predict.}
\end{figure}

To further analyze the temperature dependence, we model the bulk and edges as parallel resistors, and define an effective resistance of the edges and bulk as $R_{eff} = R_{14,23}/f$, where $f$ is the fraction of current flowing in each channel. $R_{eff}$ in our 4-terminal geometry does not depend on the contact resistance and is directly proportional to the actual resistance of the edges \cite{som}. Avoiding  the contact resistance’s effect allows us to more strongly interpret data at high temperatures (which was not the case for Ref. \cite{nowack13nmat}). $R_{eff}$  vs. temperature is presented in FIG.~\ref{fig3}. As a function of temperature, the bulk $R_{eff}$ decreases strongly with temperature, consistent with thermally activated carriers \cite{du13arxiv}. The top edge's $R_{eff}$ decreased by a factor of two over this temperature range, consistent with a constant resistance per length of the edge, but with the length along the vertical leads (compare FIG.~\ref{fig1}f) getting shorted by the bulk. This behavior is confirmed, as discussed above, by images at gate voltages with moderate bulk conduction. \cite{som}. The bottom edge is not susceptible to this effect, and $R_{eff}$of the bottom edge remained constant within the sensitivity of our analysis from 4.5 K to 32.5 K.

We have shown that the resistances of long QSH edges are unchanged up to high temperatures where the rate of inelastic backscattering should naively be varying the most. Additionally, transport measurements have shown that the resistances of long devices remain constant from 20 mK to 4 K \cite{du13arxiv}. With these two results, the backscattering mechanism in the QSH edge states of InAs/GaSb does not vary strongly in any measured temperature regime.

In contrast to this experimental result, inelastic scattering mechanisms predict temperature dependence of the edge conductivity. Generic inelastic scattering centers lead to a $T^4$ or stronger temperature dependent reduction of the conductivity \cite{strom10prl,schmidt12prl,lezmy12prb,crepin12prb}, which our data and Ref \cite{du13arxiv} firmly rule out experimentally for InAs/GaSb. Charge puddles formed by disorder can couple via tunneling \cite{vayrynen13prl} or directly to the edge states \cite{roth09sci} and both can induce backscattering. The effect of the coupling on conductivity as a function of temperature depends on the hierarchy of thermal, puddle and edge energies and the number of electrons in the puddle \cite{vayrynen13prl, vayrynen14unpub}. For odd-electron puddles formed by electric potential disorder, the Kondo effect may lead to sub-power-law temperature dependence above the Kondo temperature \cite{maciejko09prl,tanaka11prl,altshuler13prl,vayrynen14unpub}. Even in this limit, the resistivity scales as $ln^2(T)$, which is inconsistent with our observations \cite{vayrynen14unpub}. If the temperature is larger than the charging energy of a puddle, the temperature dependence may saturate \cite{vayrynen14unpub}. However, the size of charge puddles required to explain temperature-independence of the resistance down to 20 mK is of order the sample size which seems unphysical. Coupling of the electron and nuclear spins leads to non-linear IV, temperature dependence, and a predicted scattering length in InAs/GaSb \cite{delmaestro13prb} that are inconsistent with our observations \cite{som}. Weaker temperature dependence may be possible in unexplored models; however, it is difficult to understand how any inelastic scattering mechanism would lead to temperature-independent conductivity over three orders of magnitude.

Elastic processes may seem like the natural way to obtain temperature-independent conductivity, but the puzzle remains unless time-reversal symmetry is spontaneously broken (e.g. \cite{pikulin14prl}). However, the predicted critical temperature of the scattering mechanism (($\sim 10 K$), in Ref. \cite{pikulin14prl} is below the highest temperature measured here. Multi-particle scattering processes may also contribute, but are not generically temperature independent \cite{lezmy12prb}.

The experimental behavior of backscattering in HgTe and InAs/GaSb is similar: it appears above a similar device length \cite{konig07sci}, it persists to low temperatures \cite{konig07sci}, and it does not strongly vary at high temperatures \cite{nowack13nmat}. This similarity is surprising because the two materials systems differ in ways that are important for candidate scattering mechanisms. InAs/GaSb is predicted to have an order of magnitude slower Fermi velocity than HgTe \cite{liu08prl,bernevig06sci} and also a calculated Luttinger interaction parameter that implies much stronger e-e interactions \cite{maciejko09prl}. Electron interactions are assumed to be weak for many proposed backscattering mechanisms. This assumption may break down in InAs/GaSb. InAs/GaSb should exhibit stronger Rashba spin orbit coupling due to its structurally asymmetric interface, and additionally may host novel effects which arise due to the separation of hole and electron layers \cite{pikulin14prl}. Both material systems exhibit the QSHE, but how they differ in the details must be explored further experimentally and theoretically to fully understand the conditions under which the ideal QSHE breaks down.

In conclusion, we imaged current flow in InAs/GaSb quantum wells and found that the edges are more conducting than the bulk in the QSH state. These QSH edge states with high resistances imply the presence of backscattering. The backscattering mechanism does not vary at low temperatures\cite{du13arxiv} nor at high temperatures (as demonstrated in this Letter). Of the predicted inelastic scattering mechanisms, none fit the experimental observations. Elastic scattering mechanisms more intuitively fit the absence of observed temperature dependence, but single-particle backscattering is not allowed.


%
%

%



\begin{acknowledgments}
We thank R. B. Laughlin, X-L. Qi, L. Glazman, J. I. V\"{a}yrynen, and V. Cheianov for useful discussions and M. E. Huber for assistance in SQUID design and fabrication. The scanning SQUID measurements were supported by Department of Energy, Office of Basic Energy Sciences, Division of Materials Sciences and Engineering, under contract DE-AC02-76SF00515. The work at Rice University was supported by DOE-DEFG0206ER46274 and Welch C-1682.

\end{acknowledgments}

\bibliography{all_ref}

\end{document}